\documentclass{article}

\PassOptionsToPackage{numbers, compress}{natbib}
\usepackage[preprint]{arxiv}

\usepackage[utf8]{inputenc} 
\usepackage[T1]{fontenc}    
\usepackage{hyperref}       
\usepackage{url}            
\usepackage{booktabs}       
\usepackage{amsfonts}       
\usepackage{nicefrac}       
\usepackage{microtype}      
\usepackage{xcolor}         
\usepackage{physics} 
\usepackage{graphicx}
\usepackage{multirow}
\usepackage{wrapfig}

\newcommand{\PG}{\mathcal{P}}
\newcommand{\CG}{\mathcal{C}}
\newcommand{\ham}{\mathcal{H}}

\title{Classical State Preparation for Variational Quantum Algorithms via Reinforcement Learning}

\newcommand{\gino}[1]{\textcolor{black}{#1}}
\newcommand*{\newintro}{\textcolor{black}}

\author{
  \makebox[0.32\linewidth][c]{{\bf Gino Kwun}}%
  \makebox[0.32\linewidth][c]{{\bf Dhanvi Bharadwaj}}%
  \makebox[0.32\linewidth][c]{{\bf Gokul Subramanian Ravi}} \\
  Computer Science and Engineering \\
  University of Michigan \\
  Ann Arbor, MI 48109, USA \\
  \texttt{\{gkwun, dhanvib, gsravi\}@umich.edu}
}

\begin{document}

\maketitle

\begin{abstract}
Variational Quantum Algorithms (VQAs) potentially offer a pathway to practical quantum advantage, but their optimization is heavily hindered by barren plateaus and numerous local minima. While classically simulable Clifford circuits can warm-start VQAs to accelerate convergence, existing heuristic-based initialization methods struggle to scale within vast combinatorial search spaces. We address this bottleneck with CRiSP (a Clifford Reinforcement Learning agent for State Preparation), which casts Clifford prefix selection as a sequential decision-making problem. CRiSP utilizes Neural-Guided Monte Carlo Tree Search, driven by a Transformer-based policy trained via self-play, to insert learned Clifford gates before fixed parameterized rotations. This enables the construction of high-quality initial states entirely through polynomial-time classical stabilizer simulation. By integrating a curriculum learning strategy that progressively expands the search horizon, the agent efficiently scales to deep circuits. Evaluated on QAOA benchmarks of up to $22$ qubits and $1{,}370$ parameters, CRiSP outperforms state-of-the-art Clifford initialization methods by a mean of $3.19\times$ (max $45.02\times$) in average energy accuracy and $2.41\times$ (max $16.01\times$) in best-achieved energy accuracy. Assessments on VQE tasks further demonstrate the framework's robustness and generalizability.
\end{abstract}

\section{Introduction}

The potential for quantum computing to deliver a computational edge is particularly strong in domains such as chemistry~\cite{kandala2017hardware}, optimization~\cite{moll_optimization}, and machine learning~\cite{biamonte2017quantum}. 
The field is currently evolving past the era of Noisy Intermediate-Scale Quantum (NISQ) hardware~\cite{preskill2018quantum}, approaching a transition phase characterized by devices with thousands of physical qubits, augmented by advanced error mitigation and early forms of quantum error correction~\cite{Preskill_2025,google_roadmap,iroadmap_2,eft}. 
\newintro{While deep algorithms such as Shor's or Grover's~\cite{Shor_1997,grover1996fast} remain out of reach due to strict fault-tolerance requirements~\cite{O_Gorman_2017}, 
Variational Quantum Algorithms (VQAs)~\cite{cerezo2021vqa} allow shallow, hardware-native circuit designs that can help mitigate noise exposure.}
Operating as hybrid quantum-classical frameworks, VQAs employ a classical optimizer to iteratively update a parameterized quantum circuit, navigating the solution landscape to minimize a target energy expectation value ($\langle H_c \rangle$)~\cite{peruzzo2014variational,kim2023evidence,farhi2014quantum}. \newintro{Improving the efficiency of VQA execution on early fault-tolerant hardware is therefore an important near-term goal.}

\newintro{On early fault-tolerant hardware, the cost of each VQA iteration is dominated by non-Clifford gates, which carry substantial overhead~\cite{kremer2025optimizing,bharadwaj2026c}. Since every circuit evaluation is expensive, limiting the total number of VQA evaluations is critical.} However, VQA optimization landscapes are notoriously difficult to navigate, frequently forcing optimizers to evaluate thousands of expensive circuits just to make marginal progress. \newintro{A strong starting point can bypass these optimization traps and substantially reduce the number of circuit executions, which motivates classically tractable state-preparation and initialization strategies~\cite{ravi2023cafqa, bharadwaj2026spiq}.}
Among simulation-compatible gate sets, the Clifford group provides a theoretically principled discrete search space. 
\begin{wrapfigure}{r}{0.62\textwidth}
    \centering
    \includegraphics[width=\linewidth, trim={2cm 2cm 1.3cm 3.3cm}, clip]{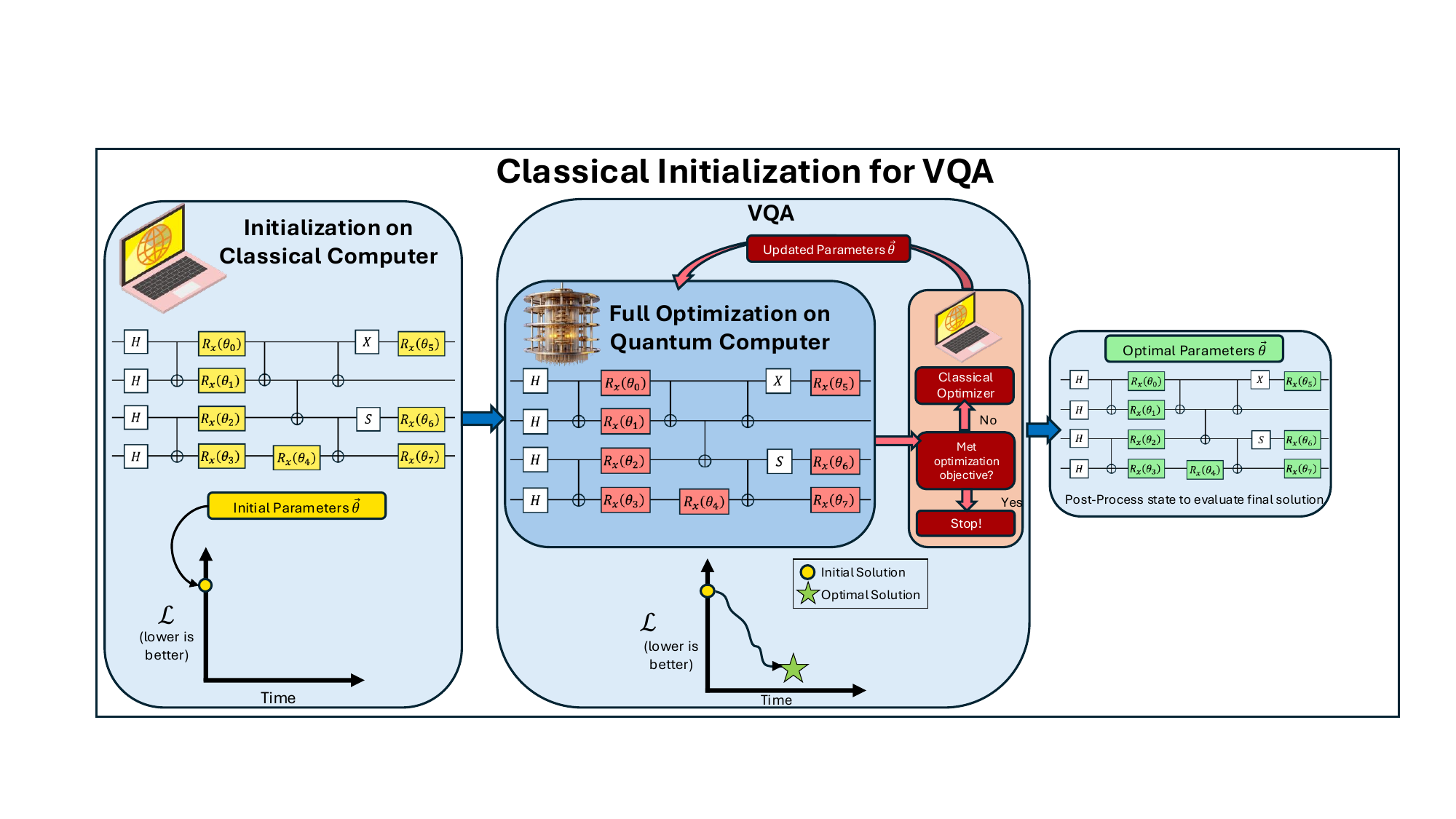}
    \caption{Classical initialization for VQA.}
    \label{fig:Initializing_VQA}
\end{wrapfigure}
Since Clifford circuits can be simulated in polynomial time classically, they enable exact computation of the initial energy expectation value without requiring quantum hardware. Existing Clifford initialization methods exploit this simulability through Bayesian or genetic-algorithm-based heuristics \cite{ravi2023cafqa,bharadwaj2026spiq}. \newintro{However, the Clifford search space grows exponentially with the number of circuit parameters, and on challenging landscapes these heuristic search techniques often struggle to navigate it. Figure~\ref{fig:Initializing_VQA} gives an overview of VQA with classical initialization.}

To overcome the scaling limitations of standard heuristics, we formulate discrete initialization as a sequential decision-making problem suited for reinforcement learning (RL)~\cite{sutton1998rl}. Specifically, we take an ansatz with fixed \newintro{rotation} gate positions and prepend a discrete Clifford gate to each rotation to act as a prefix modifier. When all continuous parameters are initialized to $\theta = 0$, the rotation gates reduce to the identity up to a global phase, \newintro{so the circuit state is determined entirely by the chosen Clifford prefixes. This state can be simulated exactly on a classical machine and serves directly as the starting point for VQA optimization. Choosing the prefixes is thus a combinatorial search task where RL has shown strong results} \cite{fawzi2022discovering,mazyavkina2021reinforcement}.

We present a Clifford Reinforcement Learning agent for State Preparation (CRiSP), a discrete-gate warm-start framework for VQAs. \newintro{Using Neural-Guided Monte Carlo Tree Search (NG-MCTS)~\cite{silver2017alphazero}, the agent searches the discrete Clifford space entirely through classical simulation, selecting sequential Clifford gates that minimize the Hamiltonian expectation value.}
At each decision step, the search is driven by a Transformer-based~\cite{vaswani2017attention} policy and value network, which encodes the current gate sequence to output prior probabilities for the next gate alongside an estimated return. \newintro{The episodic reward is the negative energy expectation value of the synthesized circuit, and the agent uses two types of replay buffers~\cite{fawzi2022discovering} to improve training efficiency. For tasks with large parameter counts and sparse energy signals, we further introduce an incremental horizon curriculum that gradually expands the search horizon during training.}

\newintro{While CRiSP applies to any VQA, we focus primarily on the Quantum Approximate Optimization Algorithm (QAOA)~\cite{farhi2014quantum}, benchmarking against SPIQ~\cite{bharadwaj2026spiq}, the state-of-the-art Clifford initialization framework, on combinatorial optimization problems including MaxCut, Knapsack, and feature selection on high-order medical datasets. We also evaluate CRiSP on the Variational Quantum Eigensolver (VQE)~\cite{peruzzo2014variational} to show that the approach generalizes beyond QAOA.}

Overall, this work makes three primary contributions. First, we formalize Clifford prefix initialization in VQA as a sequential decision-making problem over the discrete gate space, developing an RL framework that is classically simulable and avoids exponential state representations. Second, we introduce CRiSP, a Transformer-guided tree search agent trained via self-play over the Clifford prefix space, incorporating a curriculum~\cite{bengio2009curriculum} with stepwise-increasing episode length to handle long-horizon tasks. \newintro{Third, we evaluate CRiSP on QAOA benchmarks with up to $22$ qubits and $1{,}370$ circuit parameters. Given the same evaluation budget, CRiSP consistently outperforms SPIQ, improving average energy accuracy by a mean of $3.19\times$ (max $45.02\times$) and best achieved energy accuracy by a mean of $2.41\times$ (max $16.01\times$). These gains come entirely from classical computation, giving the downstream VQA a better starting point at no additional quantum cost.}

\section{Background and related work}

\paragraph{Variational quantum algorithms and optimization traps.}
VQAs are iterative hybrid quantum-classical methods, including the VQE~\cite{peruzzo2014variational} and the QAOA~\cite{farhi2014quantum}. VQE is widely used in quantum chemistry and condensed matter physics, whereas QAOA targets combinatorial optimization problems, particularly Quadratic Unconstrained Binary Optimization (QUBO). In each VQA iteration, a classical optimizer~\cite{9259985, SPSA} updates a parameterized quantum circuit (ansatz) to approximate the ground-state energy of a problem Hamiltonian~\cite{mcclean2016theory}. 
However, achieving rapid convergence is notoriously difficult. VQA optimization landscapes feature numerous local minima and barren plateaus~\cite{mcclean2018barren, Sack_2021, blekos2024review}—regions where gradients vanish exponentially as circuit size and depth increase. These traps make convergence from arbitrary initial points highly unreliable, potentially curtailing quantum benefits.

\textbf{Fault-tolerant costs and classical simulability.}
In fault-tolerant quantum computing (FTQC)~\cite{shor1996fault}, circuit cost is dominated by non-Clifford gates~\cite{kremer2025optimizing}. Arbitrary-angle rotations such as $R_x(\theta)$ and $R_z(\theta)$ must be approximately decomposed into a universal gate set, most commonly Clifford$+T$, using compilers such as Gridsynth~\cite{ross2016gridsynth} or tensor-network-based synthesizers~\cite{hao2025reducing}. Each resulting $T$ gate requires costly magic state distillation or cultivation~\cite{bravyi2005universal,gidney2024magic}, making every VQA iteration expensive. It is therefore attractive to do as much of the optimization as possible classically, before using quantum hardware.
Classical simulation, however, is efficient only in restricted regions of the state space. The Clifford group defines one such region, and it ties directly to the cost model above: when $\theta$ is a multiple of $\pi/2$, a rotation reduces to a Clifford gate and requires no $T$ gates. By the Gottesman-Knill theorem, Clifford circuits can be simulated in polynomial time and space with stabilizer tableau methods~\cite{gottesman1998heisenberg, aaronson2004improved}, so their energy expectation values can be computed exactly without quantum hardware~\cite{gidney2021stim}. Other tractable regions include low-entanglement states simulated by tensor networks~\cite{rudolph2023tensornet,ballarin2025efficient} and circuits with polynomially sized dynamical Lie algebras~\cite{goh2025lie}. These are complementary to the Clifford space, which is the focus of this work.

\paragraph{Classical VQA initialization.}
VQA performance is highly sensitive to initialization~\cite{grant2019initialization}. Early heuristics either exploit ansatz-specific structure, as in INTERP~\cite{INTERP} and TQA~\cite{Sack_2021} for QAOA, or mitigate barren plateaus without using the target Hamiltonian, as in identity-block initialization~\cite{grant2019initialization}. Clifford-based methods instead use stabilizer simulation to evaluate the target energy exactly before any quantum hardware is used. CAFQA~\cite{ravi2023cafqa} and \citet{cheng2025clifford} search the Clifford space with Bayesian optimization and simulated annealing, respectively. Clapton~\cite{seifert2024clapton} adds noise-aware transformations and a genetic algorithm, and SPIQ~\cite{bharadwaj2026spiq}, our state-of-the-art baseline, extends this approach to QAOA~\cite{farhi2014quantum}. However, these randomized or population-based heuristics struggle to scale, since the Clifford search space grows exponentially with the number of rotation gates. CRiSP instead trains a Transformer-guided tree search agent that selects a Clifford prefix for each rotation gate, so the VQA begins from a classically computed stabilizer state and refines it on quantum hardware rather than searching from scratch. Appendix~\ref{sec:additional_related_work} discusses initialization based on the other tractable regions above, as well as circuit-construction methods.

\paragraph{RL for quantum search.}
RL \cite{sutton1998rl} shows strong potential in quantum applications, from error correction \cite{sivak2026reinforcement} to circuit optimization \cite{weiden2025making}. Search-based methods like MCTS \cite{browne2012mcts} and AlphaZero \cite{silver2017alphazero} effectively navigate massive combinatorial spaces. Within the quantum domain, AlphaTensor-Quantum \cite{ruiz2025quantum} applies learned search to minimize T-counts. Most similarly, MCTS-based QAOA tuning methods~\cite{yao2022mctsqaoa,agirre2025monte} jointly optimize discrete and continuous variables or effectively restrict the search space using regular parameter patterns, but both require costly quantum circuit evaluations during search. In contrast, CRiSP performs prefix selection entirely within the classically simulable Clifford space. By training a Transformer-based policy-value network, we enable principled exploration while eliminating quantum hardware overhead during initialization. 

\section{Methods}

\begin{figure}[t]
    \centering
    \includegraphics[width=0.95\linewidth, trim={0cm 75pt 0cm 60pt}, clip] {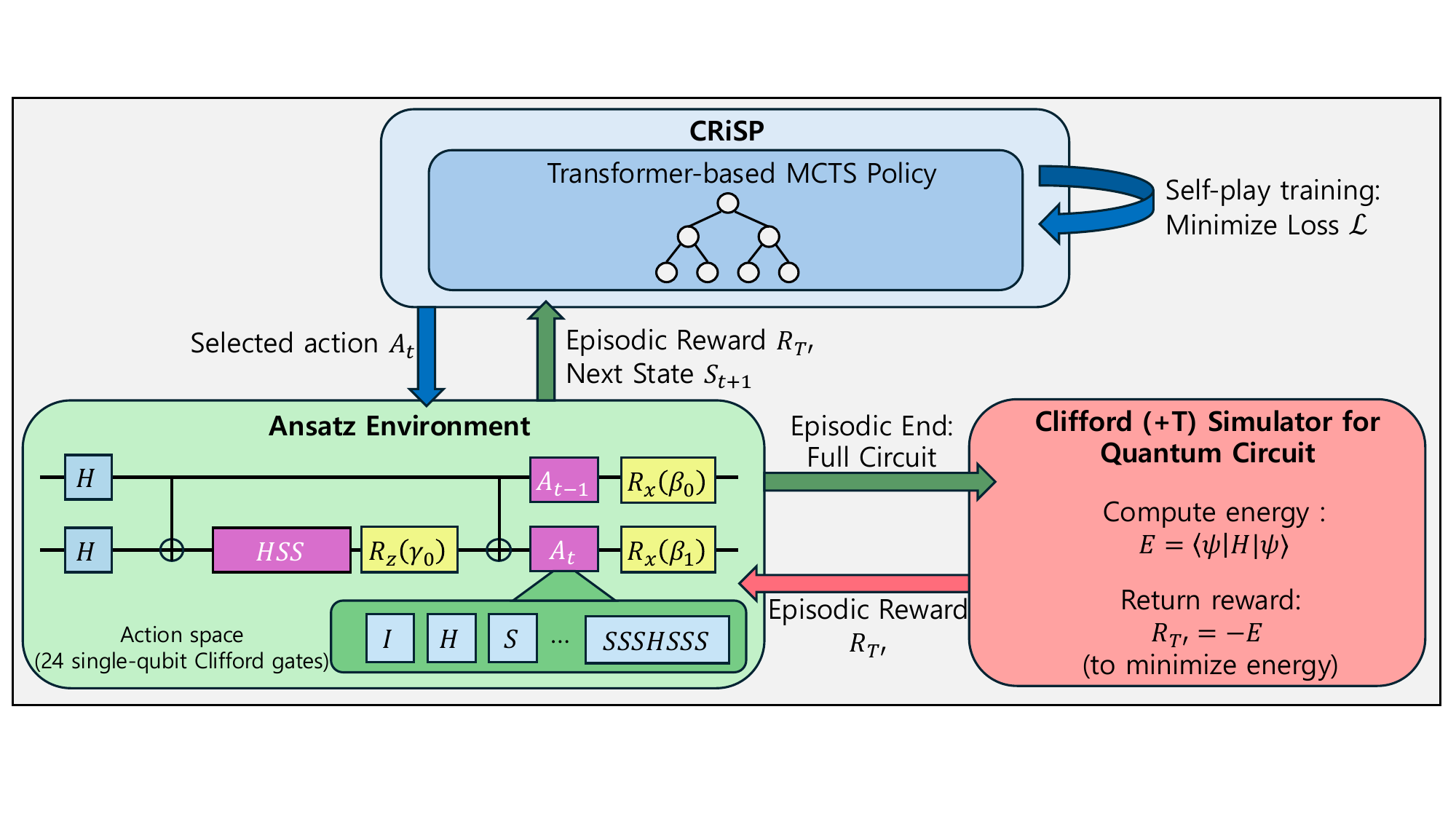}
    \caption{Overview of our proposed neural-guided framework (CRiSP) for state preparation in VQAs. The HSS gate positioned before $R_z(\gamma_0)$ represents the action $A_{t-2}$. The environment depicts a multi-angle QAOA ansatz~\cite{herrman2022maqaoa}. The circuit's final state $\ket{\psi}$ is generated by setting $\gamma_0 = \beta_0 = \beta_1 = 0$.}
    \label{fig:method_overview}
\end{figure}

In this section, we describe the configuration of the reinforcement learning (RL) environment and agent for the VQA state preparation task, as depicted in Fig.~\ref{fig:method_overview}, given a problem Hamiltonian and a fixed underlying circuit structure. We first specify the environment's state, action, and reward formulations. We then introduce CRiSP, the NG-MCTS agent that performs stepwise inference to identify the Clifford prefix configuration yielding the lowest circuit energy, alongside its self-play training procedure and a curriculum variant. Formally, given an $n$-qubit cost Hamiltonian $H_c$ 
and a fixed ansatz containing $D = \mathrm{poly}(n)$ rotational gates, such as hardware-efficient and QAOA architectures~\cite{kandala2017hardware, farhi2014quantum, cerezo2021vqa}, we cast the state preparation process as a sequential decision-making problem defined by the tuple ($\mathcal{S}, \mathcal{A}, \mathcal{T}, \mathcal{R}, \gamma$).
The agent's objective is to construct a gate sequence that prepares a quantum state $\ket{\psi}$ minimizing the expected energy $E = \langle H_c \rangle = \bra{\psi} H_c \ket{\psi}$.

\subsection{RL environment}

The \textbf{state} $s_t$ at step $t$ in the RL environment is defined as the ordered sequence of Clifford gates assigned to the first $t$ prefix positions, preceded by a \texttt{START} token.
Since the sequence length scales linearly with $D$, this representation avoids the exponential dimensionality of storing the raw quantum state vector.
Transitions are deterministic, where $s_{t+1}$ is formed by appending gate action $a_t$ to $s_t$.
An episode begins with the \texttt{START}-only state and terminates either when all gate prefix positions have been assigned ($t=D$) or when the agent reaches the maximum step count $T'$ ($t=T'$).

The \textbf{action space} comprises all 24 single-qubit Clifford gates~\cite{epstein2014investigating}, each representing a candidate prefix for the corresponding rotational gate, since the target gates are single-qubit operations.
We consider all axes regardless of the associated rotation gate's axis, opening the possibility of configurations that yield favorable energy through interactions with other circuit components.
While exhaustive search over the discrete Clifford space of $24^D$ configurations is intractable for practical system sizes, the proposed RL agent addresses this by concentrating exploration on promising regions, effectively reducing the number of nodes evaluated during tree traversal.

The \textbf{reward} $R$ is provided upon episode termination and is defined as the negative energy expectation value of the synthesized circuit under $H_c$, $ R = -E = -\langle H_c \rangle = -\bra{\psi} H_c \ket{\psi}$, where the resulting state $\ket{\psi}$ is obtained by setting all continuous parameters to $\theta = 0$.
Intermediate transitions receive zero reward, and the discount factor is set to $\gamma = 1$, reflecting the episodic task structure.
Since all gates in the synthesis procedure are Clifford, the reward can be evaluated efficiently via a classical stabilizer tableau simulator such as Stim \cite{gidney2021stim}.
Because the magnitude of $\langle H_c \rangle$ varies substantially across different Hamiltonians (from sub-unit values to over $10^4$), directly using the raw reward as a training target introduces significant gradient instability.
To address this, we normalize raw reward targets in the buffers using a running mean $\mu$ and standard deviation $\sigma$ as $R_\text{norm} = \frac{R - \max(\mu, 0)}{\sigma + \epsilon} - 1$, where $\epsilon$ is a small numerical offset.
The $-1$ term penalizes $R=0$ cases, encouraging the agent to seek negative energies (positive rewards).
The normalized reward is subsequently clipped to $[-1, 1]$ to suppress instability from outlier values.
This adaptive rescaling continuously rebalances the reward range, sustaining active search for higher-return configurations.

\subsection{The CRiSP framework}

We now introduce CRiSP, a Clifford Reinforcement Learning agent for State Preparation in VQA, which utilizes Neural-Guided MCTS (NG-MCTS) to discover the best gate action in each step.

\paragraph{NG-MCTS procedure.}
At each step within an episode, CRiSP selects the next gate action via MCTS guided by a neural network that estimates the expected return of a given state and the prior action probabilities used in simulation \cite{silver2017alphazero}.
The search proceeds through three stages, namely selection, expansion, and backpropagation.
In the \textbf{selection} phase, the agent traverses the tree by iteratively choosing actions according to the following upper confidence bound (UCB) criterion \cite{silver2017alphazero} when in state $s$,

\begin{equation}
    a_{\text{next}} = \arg\max_{a} \left(Q(s, a) + c_\text{puct} \cdot 
    P(s, a) \cdot \frac{\sqrt{N(s)}}{1 + N(s, a)}\right),
\end{equation}

where $Q(s, a)$ is the empirical expected state-action value, $P(s, a)$ is the prior probability predicted by the network, and $N(s)$ and $N(s, a)$ denote the visit counts of the parent and child nodes, respectively.
The coefficient $c_\text{puct}$ controls exploration within each simulation, where higher values encourage broader search.
To further promote exploration, Dirichlet noise \cite{silver2017alphazero} is injected into the root node's prior probabilities at each episode step.
Upon reaching a leaf node, the tree undergoes \textbf{expansion}, where new child nodes are added and the neural network evaluates the leaf state value along with prior probabilities for its children.
If the leaf is terminal, no expansion occurs and the estimated value is passed directly to the next procedure.
During \textbf{backpropagation}, this value is propagated upward to the root, updating the mean state-action values and visit counts of all non-terminal nodes along the traversed path.
After collecting $m$ simulation results per step, the agent samples the next gate proportionally to the visit counts of the child nodes.
To balance exploration and exploitation and facilitate convergence, step-wise decay is applied to the temperature $\tau$ when sampling the next action \cite{silver2017alphazero}.
Between consecutive steps within an episode, the search tree is reused by re-rooting at the selected action node, improving simulation quality.

\paragraph{Neural network architecture.}

The policy and value networks directing MCTS are composed of four modules: an embedding layer, a shared network, a policy head, and a value head.
Given the current gate sequence and the problem Hamiltonian, the embedding layer maps each input token into a dense vector representation augmented with sinusoidal positional encodings \cite{vaswani2017attention}.
Hamiltonian terms are encoded via one-hot representations of their Pauli structure, with coefficients normalized to balance the relative contribution of each term and prevent numerical instability from outliers. The Hamiltonian encoding is added to the state embedding with a learnable scaling value, forming the input embedding to the shared network.
The shared network is a multi-head Transformer encoder \cite{vaswani2017attention}, which captures sequential dependencies among gate tokens in the current prefix.
From the final hidden representation, two separate heads produce distinct outputs, where the policy head applies a softmax over the action space to generate prior probabilities for the next gate and the value head computes a scalar estimate of the expected return.
To improve value discrimination among states at the same prefix depth, the value head receives the concatenation of the final Transformer output with the embedding of the most recently added gate token as input.
To bound memory consumption, the model processes only the last $c_\text{ctx}$ tokens of the sequence, where $c_\text{ctx}$ is a hyperparameter.
Layer normalization~\cite{ba2016layernorm} and the GELU activation function~\cite{hendrycks2016gelu} are employed to stabilize forward-pass signal propagation.

\paragraph{Agent training.}

The neural network is trained through experience replay to progressively improve simulation quality.
At each episode step, a sample consisting of the current gate prefix, the MCTS-derived policy target, and the terminal episode reward is stored in a replay buffer.
In addition, a compact best-game buffer~\cite{fawzi2022discovering} retains high-reward trajectories to facilitate convergence toward promising search regions, as well as to prevent zero or negative signals from dominating the learning process.
In each training round, the network is trained for $K$ epochs using the AdamW optimizer~\cite{loshchilov2017adamw} and a linear learning rate warmup~\cite{vaswani2017attention}, with each epoch sampling mini-batches from the replay buffer mixed with a fixed proportion of best-game samples once a sufficient number of episodes have been collected.
The total training loss $\mathcal{L}$ combines policy and value objectives as: $\mathcal{L} = \mathcal{L}_\text{p} + c_v \mathcal{L}_\text{v}$, where $\mathcal{L}_\text{p}$ is the cross-entropy loss \cite{mao2023cross} between the current policy output and the MCTS-derived target distribution and $\mathcal{L}_\text{v}$ is the Huber loss \cite{huber1992robust} between the predicted and realized returns, balanced by the coefficient $c_v$.
Throughout training, the policy is periodically evaluated by greedily selecting the highest-probability action at each step with $m_\text{eval}$ MCTS simulations and reduced Dirichlet noise, tracking whether the learned model recovers the high-reward configurations found during training.

\subsection{Curriculum learning for long-horizon and sparse-reward tasks} \label{subsec:curriculum}

To address tasks with extremely large maximum episode lengths and sparse-reward signals by progressively guiding the agent through increasingly complex search horizons, we devise a curriculum variant of CRiSP, which we call the incremental horizon curriculum. 
Inspired by~\citet{bengio2009curriculum} and the role of horizon length in offline RL scalability~\cite{park2025horizon}, this strategy consists of three stages with monotonically increasing maximum step counts.
This curriculum is made possible by the early-termination property of the target RL environment, which enables the episodic evaluation at any circuit construction step by treating all unassigned prefix positions as identity ($I$) gates.
Specifically, the agent operates with $25\%$ of the full episode length ($0.25 T'$) until $25\%$ of total training episodes have elapsed, then expands to $50\%$ ($0.5 T'$) until the midpoint of training, and finally opens the complete horizon ($T'$) for the remainder.
Although this naive staged schedule may be suboptimal for some cases, empirical results confirm that the strategy is particularly effective on benchmarks with large search spaces and limited simulation budgets, where concentrating early training on shorter, tractable sequences accelerates identification of promising partial configurations.

\section{Evaluation}

In this section, we evaluate CRiSP on multiple standard VQA benchmarks of practical interest. We first detail the experimental methodology (Sec.~\ref{sec:method}) and present the QAOA benchmark results (Sec.~\ref{sec:qaoa}). Next, we perform downstream VQA optimization to investigate the effect of initialization (Sec.~\ref{sec:post_init_results}). Finally, we present exploratory results applying CRiSP to the VQE domain to demonstrate the broader generalizability of our framework (Sec.~\ref{sec:vqe}). Implementation details are provided in Appendix~\ref{sec:further_exp_detail}, and additional experiments can be found in Appendix~\ref{sec:extra_experiment}.

\subsection{Methodology}
\label{sec:method}

\paragraph{QAOA benchmarks.} To evaluate CRiSP, we consider three classes of combinatorial optimization problems, with Hamiltonian derivations following~\cite{bharadwaj2026spiq}. First, we adopt MaxCut on weighted complete graphs as our primary Quadratic Unconstrained Binary Optimization (QUBO) benchmark, providing highly entangled and globally coupled optimization landscapes. Second, for Polynomial Constrained Binary Optimization (PCBO) evaluation, we utilize the Knapsack problem formulated as a Quadratic Unconstrained Binary Optimization (QUBO) Hamiltonian \cite{qiskit23}. Third, we evaluate a biomedical high-order feature selection task \cite{cancer_genome}, comprising the ``MedicalDense'' and ``MedicalSparse'' benchmarks. We specifically include the latter because its extreme sparsity induces severe barren plateaus, presenting a critical challenge for VQA optimization. Detailed formulations for these Hamiltonians are deferred to Appendix~\ref{subsec:qaoa_problem_derivation}.

\paragraph{Comparative baseline.}
To benchmark the quality of the classically-prepared state for these QAOA tasks, we compare CRiSP against SPIQ \cite{bharadwaj2026spiq}, a state-of-the-art Clifford initialization method leveraging a genetic algorithm (GA).
All QAOA experiments are conducted in multi-angle QAOA (ma-QAOA) \cite{herrman2022maqaoa} circuits with $p = 1$, where all parameterized rotations are relaxed, thereby allowing each parameter to have its respective gate prefix.

\paragraph{Metrics.} To enable consistent performance comparison across all tasks with substantially different energy scales, we report an accuracy metric similar to that employed in~\citet{bharadwaj2026spiq}, defined as $\text{Accuracy} = {E_\text{best}}/{E_\text{opt}}$,
where $E_\text{best}$ denotes the minimum energy achieved by a given method and $E_\text{opt}$ is the ground-state energy of the corresponding Hamiltonian. 
Across experiments, we employ three random seeds and report the mean and standard deviation of the results.

\begin{figure}[t]
    \centering
    \includegraphics[width=\linewidth]{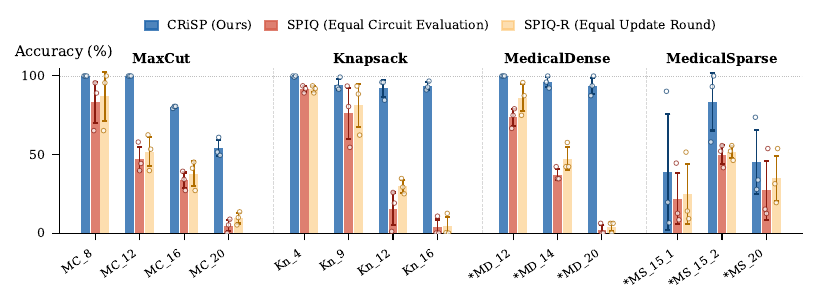}
    \caption{Classical initialization accuracy for QAOA benchmarks. Asterisks (*) indicate tasks using the incremental horizon curriculum. The bar reflects the average accuracy, the error bar represents the standard deviation (std), and the points show results from three independent runs. }
    \label{fig:qaoa_results}
\end{figure}

\subsection{QAOA results}
\label{sec:qaoa}
To assess the CRiSP's performance, we compare the proposed agent against SPIQ~\cite{bharadwaj2026spiq} on combinatorial optimization problem suites formulated as Hamiltonians. These include MaxCut on weighted complete graphs, the Knapsack problem, and a biomedical high-order interaction task~\cite{cancer_genome}. Both methods employ QAOA circuit architectures to search for the ground state of each respective Hamiltonian. While CRiSP is broadly applicable to arbitrary Parameterized Quantum Circuits (PQCs), we adopt a ma-QAOA~\cite{herrman2022maqaoa} framework to ensure a fair comparison, since SPIQ is specifically tailored to such architectures. Following~\citet{bharadwaj2026spiq}, we set the population size of each GA generation to 100. However, we adopt a single-start configuration, as our empirical testing showed that it generally yields superior performance to the original multi-start setting under an equivalent computational budget. 

Since both frameworks rely on iterative refinement processes to improve solutions using accumulated performance data, we evaluate SPIQ under two comparative conditions. Our primary baseline matches SPIQ to CRiSP by the total number of circuit evaluations. As a supplementary baseline, we also report the performance of SPIQ-R with the number of update iterations matched to CRiSP (i.e., SPIQ generations are equated to CRiSP's training rounds, where each CRiSP round comprises episode collection from 30 workers followed by $K$ epochs of network updates). Furthermore, for the medical benchmarks~\cite{cancer_genome}---featuring either extreme episode lengths up to $1{,}370$ parameters (MedicalDense) or exceptionally sparse reward landscapes (MedicalSparse)---we report CRiSP results utilizing the incremental horizon curriculum detailed in Section~\ref{subsec:curriculum}.

Figure~\ref{fig:qaoa_results} summarizes the QAOA benchmark results, demonstrating that CRiSP consistently achieves higher energy accuracy than the competing approaches across all tasks (see Appendix~\ref{sec:numerical_results} for numerical details). With an overall mean accuracy of $83.7\%$ across 14 distinct tasks, CRiSP outperforms the evaluation-matched SPIQ baseline by a geometric mean (GeoMean) of $3.19\times$ in average energy accuracy and $2.41\times$ in best-achieved energy accuracy. Maximum improvements peak at $45.02\times$ and $16.01\times$, respectively, on the MedicalDense\_20 task. Against SPIQ-R, CRiSP achieves GeoMean improvements of $2.51\times$ (average) and $2.13\times$ (best), with maximum gains of $22.51\times$ (average) and $16.01\times$ (best), again on the MedicalDense\_20 benchmark.

A categorical breakdown further underscores the robustness of the proposed framework. Compared to the evaluation-matched baseline, CRiSP achieves GeoMean improvements of $2.89\times$ (average) and $2.25\times$ (best) on MaxCut, $3.81\times$ (average) and $2.48\times$ (best) on Knapsack, and $5.41\times$ (average) and $3.63\times$ (best) on MedicalDense tasks. Notably, even on MedicalSparse tasks---where extreme landscape sparsity makes it inherently difficult for either method to discover non-zero reward configurations---CRiSP still realizes GeoMean advantages of $1.70\times$ (average) and $1.71\times$ (best) higher accuracy than SPIQ across all instances. 

For the MedicalSparse tasks specifically, we observe a wider accuracy distribution among different random seeds, with certain CRiSP instances occasionally underperforming SPIQ. While consistent outperformance across all seeds is ideal, identifying even a single high-quality solution holds immense practical value in realistic quantum computing use cases, as it drastically reduces subsequent quantum resource expenditures. Consequently, the fact that the best-performing seeds consistently originate from CRiSP is of greater practical utility than strictly outperforming the baseline on average. We hypothesize that performance on these sparse tasks is primarily bottlenecked by the inherent difficulty of initial exploration rather than the search policy itself. This is evidenced by the substantially larger performance margins that CRiSP demonstrates on denser benchmarks, where non-zero reward signals become more readily available to guide learning.

\begin{figure}[t]
    \centering
    \includegraphics[width=\linewidth]{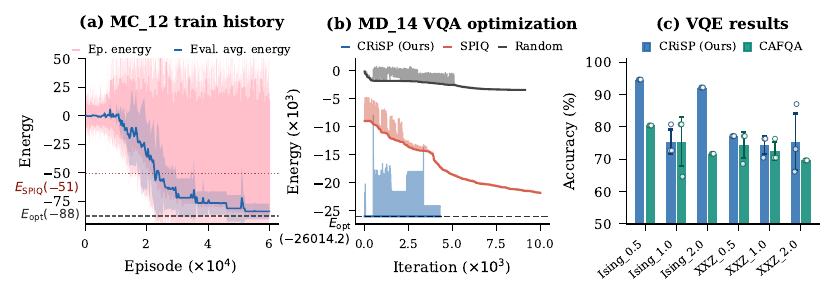}
    \caption{(a) MaxCut\_12 training history. $E_\text{opt}$ and $E_\text{SPIQ}$ denote the ground-state and the best SPIQ energies, respectively. The shaded areas denote std across three seeds. (b) VQA optimization trajectories of the MedicalDense\_14 instance, starting from CRiSP, SPIQ, and random initialization. Solid lines show the best energy found so far for each method, while the faint lines trace the raw per-iteration energy values. (c) Initialization accuracy for VQE tasks. The bar indicates the average accuracy, the error bar displays std, and the points correspond to results from three independent runs. }
    \label{fig:energy_cont_opt_vqe}
\end{figure}

Finally, the performance gap between the two methods notably widens as problem dimensionality scales. While SPIQ degrades significantly on larger instances, CRiSP exhibits strong scalability, maintaining average energy accuracies above $92\%$ throughout all Knapsack and MedicalDense benchmarks. This consistent improvement is further corroborated by the convergence behavior illustrated in Fig.~\ref{fig:energy_cont_opt_vqe}(a), which confirms that CRiSP progressively and reliably converges toward low-energy (high-return) configurations given sufficient training budgets.

\subsection{Downstream VQA optimization results}
\label{sec:post_init_results}

To evaluate whether better initialization translates into better VQA outcomes, we run downstream optimization from CRiSP, SPIQ, and random initializations under an identical budget of $10^4$ quantum circuit evaluations.
Fig.~\ref{fig:energy_cont_opt_vqe}(b) shows the optimization trajectories on the MedicalDense\_14 instance using the Constrained Optimization BY Linear Approximations (COBYLA) optimizer~\cite{Zhang_2023}, tracking the sample with the best final energy for each method.

CRiSP reaches a substantially lower final energy than SPIQ and random initialization.
In fact, CRiSP's initial energy is already lower than the final energy of either baseline, so its advantage does not depend on additional quantum evaluations.
Because this starting point is already highly accurate, the further gain from optimization is limited; it is nonzero but less visible at the scale of the figure.
The same trend holds across all benchmark categories (Appendix~\ref{sec:extra_post_init_results}): CRiSP reaches a lower final energy, and in cases where the baselines come close, they require many more quantum evaluations to do so.
These results indicate that a better initialization leads to better downstream performance by starting the optimization in a favorable region of the landscape.
This is consistent with prior work showing that VQA landscapes contain many poor local minima~\cite{anschuetz2022quantum} and that good initialization substantially improves QAOA performance~\cite{INTERP}.
How much of this benefit is realized also depends on the choice of continuous optimizer; to our knowledge, this remains an open question, which we view as orthogonal to finding a good initialization.

\subsection{Framework generalization to VQE}
\label{sec:vqe}

To investigate the applicability of CRiSP beyond the QAOA domain, we preliminarily evaluate the framework on VQE benchmarks using two physical models: the one-dimensional Transverse-Field Ising Model (TFIM)~\cite{pfeuty1970one} and the XXZ model~\cite{heisenberg}, a specific variant of the Heisenberg model. 
Detailed Hamiltonian formulations for these models are provided in Section~\ref{subsec:vqe_problem_derivation}.
Both models are evaluated at $N=10$ across three coupling strengths, $J \in \{0.5, 1.0, 2.0\}$.

We compare CRiSP against a scalable variant of CAFQA~\cite{ravi2023cafqa} (implemented via a GA, as employed in Clapton~\cite{seifert2024clapton}), a Clifford initialization method designed for VQE, under equal circuit evaluation budgets. Both approaches utilize a circular hardware-efficient ansatz~\cite{kandala2017hardware} with one repetition, resulting in $40$ parameters per circuit. Given the exploratory nature of this experiment, we adopt the closest QAOA benchmark configuration of comparable problem size without VQE-tailored algorithmic setup, performing a grid search over the per-round training epochs $K \in \{12, 16, 20, 24\}$ and the value coefficient $c_v \in \{1, 2\}$.

As presented in Fig.~\ref{fig:energy_cont_opt_vqe}(c), CRiSP achieves comparable or higher accuracy than CAFQA across all benchmarks, with GeoMeans of $1.1\times$ in average accuracy and $1.11\times$ in best-achieved accuracy (see Appendix~\ref{sec:numerical_results} for details). On Ising\_0.5 and Ising\_2.0, CRiSP consistently outperforms CAFQA by $1.18\times$ and $1.29\times$, respectively. On the XXZ models, although the variance across seeds increases with the coupling strength $J$, CRiSP successfully identifies the best circuit configurations across all benchmarks, some of which remain unreachable by CAFQA. 
While additional hyperparameter tuning would likely further enhance performance, these preliminary results indicate that CRiSP generalizes to VQE settings without requiring task-specific modifications, supporting its potential applicability to a broader class of VQA problems.

\section{Conclusions, limitations, and future outlook}
\label{sec:conclusion}

We introduced CRiSP, a reinforcement learning framework for classically simulable state preparation that addresses the resource cost of VQAs on early fault-tolerant hardware. By casting Clifford initialization as a sequential decision-making problem, CRiSP uses Transformer-guided Monte Carlo Tree Search to navigate the exponentially large Clifford search space. Across QAOA benchmarks with up to $22$ qubits and $1{,}370$ parameters, as well as VQE, CRiSP consistently outperforms the state-of-the-art heuristic baseline, improving average energy accuracy by a mean of $3.19\times$ under the same circuit evaluation budget.

Our work has some limitations. First, although the empirical results in this paper and in prior work~\cite{bharadwaj2026spiq} clearly show that lower initial energies lead to better results after continuous optimization, there is no formal guarantee that this holds in general. Second, as with many near-term algorithms, VQAs are not guaranteed to achieve practical advantage over advanced classical methods, though their noise resilience and flexibility justify continued study. Third, while each stabilizer simulation remains polynomial and our curriculum handles large parameter counts, the Clifford search space still grows exponentially, so training cost increases with system size.

Building upon these insights, we identify five highly promising directions for future research:

\textbf{Landscape-aware initialization:} Reward signals could incorporate landscape properties such as gradient or Hessian magnitudes, guiding the agent toward low-energy states in trainable regions and accelerating subsequent optimization.

\textbf{Near-Clifford state preparation:} Allowing a small budget of non-Clifford gates (e.g., $T$ gates) expands the reachable state space, at a simulation cost that grows exponentially only in the number of non-Clifford gates. Extending CRiSP to this action space could yield higher-quality initial states.

\textbf{Zero-shot or few-shot transfer:} A model trained on a set of problem instances that generalizes to unseen ones would reduce per-instance training cost.

\textbf{Generalization beyond VQAs:} Replacing the energy objective with state or unitary fidelity would adapt our sequential construction framework to approximate state preparation and circuit compilation, for example to reduce $T$-count or multi-qubit gate count.

\textbf{Quantum-in-the-loop exploration:} Evaluating candidates on quantum hardware would let the search reach non-Clifford states that cannot be simulated classically. Given the cost of quantum execution, this calls for sample-efficient RL in the limited-shot regime.

\begin{ack}
This material is based upon work supported by the U.S. Department of Energy, Office of Science, Office of Advanced Scientific Computing Research, Accelerated Research in Quantum Computing under Award Number DE-SC0025633. This research used resources of the National Energy Research Scientific Computing Center, a DOE Office of Science User Facility supported by the Office of Science of the U.S. Department of Energy under Contract No. DE-AC02-05CH11231 using NERSC award NERSC DDR-ERCAP0035341. This research was, in part, funded by the U.S. Government. The views and conclusions contained in this document are those of the authors and should not be interpreted as representing the official policies, either expressed or implied, of the U.S. Government.

\end{ack}

\bibliographystyle{plainnat}
\bibliography{gino_ref,gokul_ref,dhanvi_ref}

\appendix

\newpage
\section*{Appendices}

\section{Additional related work}
\label{sec:additional_related_work}

\gino{Beyond Clifford-based initialization, several other strategies aim to give VQAs a better starting point, each targeting a different regime. 
Tensor-network-based pretraining~\cite{rudolph2023tensornet} represents the circuit state with a truncated bond dimension. 
This allows arbitrary non-Clifford states but caps how much entanglement the state can carry, which is the opposite trade-off from Clifford-based methods: those are restricted to stabilizer states but place no cap on entanglement. 
The entanglement cap matters for QAOA in particular. Even when the optimal solution is a product state, the QAOA circuit can pass through a volume-law entanglement barrier in the middle of its evolution~\cite{dupont2022entanglement}. 
An entanglement-limited representation cannot faithfully capture these intermediate states, which can limit the quality of the initial state it produces.}

\gino{Lie-algebraic simulation and initialization techniques~\cite{goh2025lie} take a different route. 
When the dynamical Lie algebra generated by the ansatz is polynomially sized and contains the observable, expectation values can be computed efficiently on a classical computer. 
In this regime, however, the entire variational problem becomes classically tractable, so these methods effectively replace the quantum algorithm rather than warm-start it. Outside this regime, they do not apply.}

\gino{Adaptive ansatz approaches such as ADAPT-VQE~\cite{adaptvqe} and generative circuit synthesis models such as GQE~\cite{nakaji2024generative} instead search over the circuit structure itself rather than warm-starting a fixed ansatz. 
ADAPT-VQE grows the ansatz one operator at a time based on measured energy gradients, while GQE trains a generative model to propose circuits scored by their energies. 
Both typically rely on quantum hardware evaluations during the search, which incurs the same per-evaluation cost that motivates classical initialization in the first place. 
In contrast, CRiSP warm-starts a fixed ansatz and performs its entire search classically via stabilizer simulation~\cite{gidney2021stim}, so no quantum evaluations are spent before the VQA begins.}

\section{Experimental details}
\label{sec:further_exp_detail}

MCTS trajectory collection is distributed across 30 parallel workers, forming 30 episodes per training round.
The training mini-batch size is set to 512 or 1024 depending on the problem size.
We employ a two-stage simulation schedule, in which the agent begins with $m = 50$ simulations per step during the warmup period, which is the first $10\%$ of the total episodes, to allow rapid exploration, then transitions to a standard phase with $m=100$ simulations to enhance search quality.
The evaluation simulation count $m_\text{eval}$ is set equal to that of the standard phase.
The model is trained using AdamW~\cite{loshchilov2017adamw} with a peak learning rate of $10^{-4}$ and a linear learning rate warmup \cite{vaswani2017attention} during the initial stage with a reduced simulation count.
To facilitate convergence, a step-wise exponential temperature decay \cite{silver2017alphazero} is applied to $\tau$ after the warmup episodes, starting from $\tau = 1.0$ at step $25$ until a horizon-dependent threshold is met, after which $\tau$ is held at $0.75$.
When the incremental horizon curriculum is applied, we use reduced $c_{\text{puct}}$ to enhance convergence on large tasks, while a grid search over $c_{\text{puct}} \in \{1, 1.5\}$ is performed in sparse-reward environments.
To encourage exploration upon each horizon expansion, the temperature is temporarily set to $\tau=1.2$ for the subsequent $5\%$ of episodes, applied only to gate selection steps within the newly expanded portion of the search horizon.
Depending on the problem size, hyperparameters including simulation counts, training episodes, epochs, and learning rates are adjusted based on resource constraints.

The implementation uses Qiskit 2.3.0~\cite{qiskit23} for general quantum circuit operations, Pytorch 2.5.1~\cite{paszke2019pytorch} for model training, Stim 1.15.0~\cite{gidney2021stim} for classical stabilizer tableau simulation to compute the expected energy of the synthesized quantum circuits, NumPy 2.3.3~\cite{harris2020array} for numerical computation, and rustworkx 0.17.1~\cite{treinish2021rustworkx} for graph operations, all running on Python 3.12.
All experiments are conducted on a single NVIDIA A100 GPU with 32 cores of a 64-core AMD EPYC 7763 CPU and up to 64GB memory.
Execution time ranges from 10 hours to 80 hours, depending on the simulation scale and budget.
Hyperparameters are reported in Table~\ref{tab:hyperparams}.
For VQE experiments, the values of $K$ and $c_v$ are listed in Table~\ref{tab:vqe_hyperparams}, with all remaining hyperparameters set to MaxCut\_8 configuration in Table~\ref{tab:hyperparams}.

\begin{table}[ht]
\centering
\caption{Hyperparameters for CRiSP training and architecture. When indicating benchmarks, MC, Kn, MD, and MS refer to MaxCut, Knapsack, MedicalDense, and MedicalSparse, respectively.}
\label{tab:hyperparams}
\scriptsize
\begin{tabular}{llc}
\toprule
\textbf{Category} & \textbf{Hyperparameter} & \textbf{Value} \\ \midrule
\multirow{8}{*}{Architecture} & Transformer layers / heads & 2 / 4 \\
 & Transformer input / FF dimension & 128 / 256 \\
 & State / Positional embedding dimension & 32 / 96 \\
 & Positional embedding / size & Sinusoidal~\cite{vaswani2017attention} \\
 & Hamiltonian embedding size & [256, 128] \\
 & Policy / Value head size & [128, 128, 24] / [256, 128, 64, 1] \\
 & Context length ($c_{\text{ctx}}$) & 64 \\
 & Activation & GELU~\cite{hendrycks2016gelu}\\
 & Normalization & LayerNorm~\cite{ba2016layernorm} \\ \midrule
 
\multirow{5}{*}{MCTS} & $c_{\text{puct}}$ & 1.0 (default), 0.8 (MD\_12/14/20), 1.5 (MS\_15\_1) \\
 & Dirichlet noise ($\alpha_D, \epsilon_D$) & 0.15, 0.2 (Eval: 0.15, 0.1) \\
 & Warmup / Standard simulations ($m$) & 50 / 100 (default), 30 / 50 (MC\_20, Kn\_16, MD\_12/14/20) \\
 & Parallel workers & 30 \\
 & Initial / Decayed temperature $\tau$ & 1.0 / 0.75 \\ \midrule
 
\multirow{13}{*}{Optimization} & \multirow{2}{*}{Total episodes} & $6\times10^4$ (default), $3\times10^4$ (MC\_20, Kn\_12/16, MD\_12), \\
& & $1.5\times10^4$ (MD\_14), $6\times10^3$ (MD\_20)\\
& \multirow{2}{*}{Training epochs ($K$)} & 16 (default), 12 (MC\_8), 24 (MC\_20, Kn\_9/12/16), \\
& & 32 (MD\_12/14), 40 (MD\_20) \\
& Optimizer & AdamW~\cite{loshchilov2017adamw} \\
 & Peak learning rate & $10^{-4}$ (default), $3 \times 10^{-4}$ (MD\_20) \\
 & Weight decay & $10^{-6}$ \\
 & Value coefficient ($c_v$) & 2.0 \\
 & Mini-batch size & 1024 (default), 512 (MC\_8/12, Kn\_4, MS\_15\_1/15\_2) \\
 & Gradient clipping & 1.0 \\
 & \multirow{2}{*}{Replay buffer size} & $5\times10^5$ (default), $10^5$ (MC\_20, Kn\_9/12/16, MD\_12, MS\_20), \\
 & & $1.5\times10^5$ (MD\_14), $4\times10^5$ (MD\_20) \\
 & \multirow{3}{*}{Best game buffer size} & $10^3$ (default), $2\times10^3$ (Kn\_9/12), \\
 & & $3\times10^3$ (MC\_20, Kn\_16, MD\_12), $5\times10^3$ (MD\_14), \\
 & & $10^4$ (MD\_20) \\
 & Training sample ratio & (0, 0) (initial), (1.0, 0) (after 300 episodes), \\
 & (replay, best game) & (0.9, 0.1) (after 600 episodes) \\
\bottomrule
\end{tabular}
\end{table}

\begin{table}[ht]
\centering
\caption{Hyperparameters for VQE training across different tasks. }
\label{tab:vqe_hyperparams}
\small
\begin{tabular}{@{}lcc@{}}
\toprule
\textbf{Task} & \textbf{Training epochs ($K$)} & \textbf{Value coefficient ($c_v$)} \\ \midrule
Ising\_0.5 & 12 & 1.0 \\
Ising\_1.0 & 16 & 2.0 \\
Ising\_2.0 & 24 & 2.0 \\
XXZ\_0.5   & 16 & 1.0 \\
XXZ\_1.0  & 12 & 1.0 \\
XXZ\_2.0  & 16 & 1.0 \\ \bottomrule
\end{tabular}
\end{table}

\section{Additional experiment results}
\label{sec:extra_experiment}

\begin{figure}[t]
    \centering
    \includegraphics[width=1.0\textwidth]{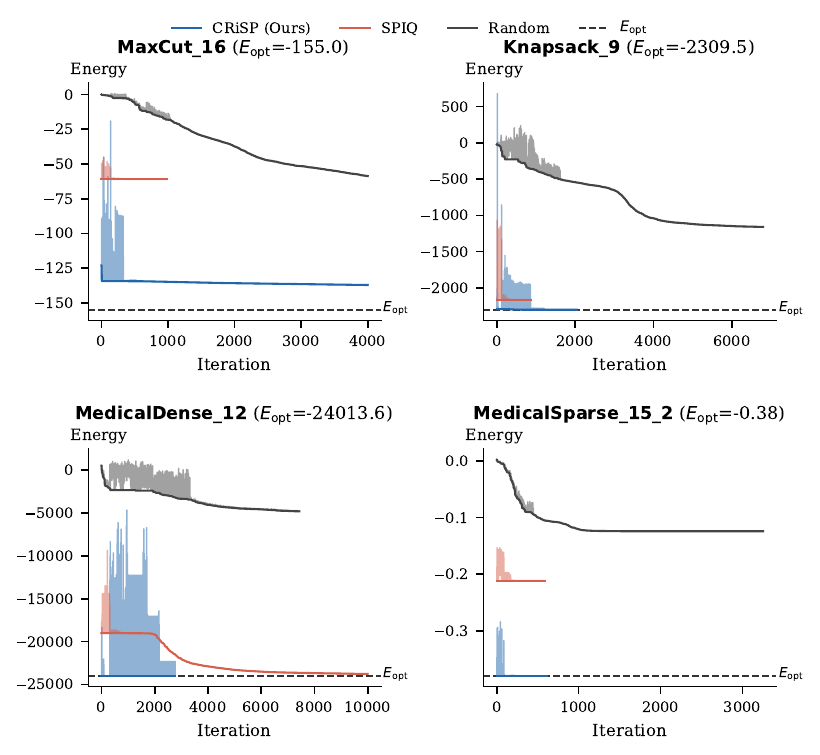}
    \caption{ 
    Downstream VQA optimization results, comparing random initialization, SPIQ, and CRiSP as starting points for continuous COBYLA optimization~\cite{Zhang_2023}. Circuit evaluation budgets are matched within an instance, set on a similar scale to the budgets used in \citet{bharadwaj2026spiq}. Solid lines show the best energy found so far for each method, while the faint lines trace the raw per-iteration energy values. Lower energy with fewer iterations is better.
    }
    \label{fig:extra_cont_opt_result}
\end{figure}

\subsection{Additional downstream VQA optimization results}
\label{sec:extra_post_init_results}

Figure~\ref{fig:extra_cont_opt_result} presents additional downstream optimization results for MaxCut\_16, Knapsack\_9, MedicalDense\_12, and MedicalSparse\_15\_2, starting from CRiSP, SPIQ~\cite{bharadwaj2026spiq}, and random initializations.
All methods receive the same quantum circuit evaluation budget for a given instance: $4{,}000$ for MaxCut\_16 and $10^4$ for the others.
Across all four instances, CRiSP achieves a lower final energy than both baselines.
In MedicalDense\_12, SPIQ approaches CRiSP's final energy, but it requires substantially more quantum evaluations to do so.

For Knapsack\_9 and MedicalDense\_12, CRiSP's initialization is already highly accurate, so the subsequent optimization gain (up to an order of magnitude lower error) is not clearly visible at the scale of the figure.
For MedicalSparse\_15\_2, optimization does not improve the CRiSP result, and the same holds for SPIQ.
This may be due to a flat landscape around the initial point, such as a barren plateau, and highlights room for further research on post-initialization optimization.
Even so, CRiSP's energy remains substantially lower than SPIQ's energy.

\begin{table}[h]
\centering
\caption{
    Time-matched comparison results on MaxCut\_12, MaxCut\_20, Knapsack\_16, and MedicalDense\_20. SPIQ-T denotes a variant of SPIQ~\cite{bharadwaj2026spiq} whose wall-time is matched to the average time CRiSP took to reach its best solution. $E_{\text{opt}}$ indicates the optimal energy. Accuracy results are reported as mean $\pm$ standard deviation (std) across three independent runs, where higher is better. Bold values denote the best average performance for each task.
}
\label{tab:time_matched_results}
\begin{tabular}{lrrr}
\toprule
Task & $E_\text{opt}$ & CRiSP (Ours) (\%) & SPIQ-T (\%) \\
\midrule
MaxCut\_12 & $-88$ & $\textbf{100} \pm 0$ & $52.27 \pm 11.97$\\
MaxCut\_20 & $-192$ & $\textbf{53.99} \pm 6.1$ & $40.63 \pm 6.53$ \\
Knapsack\_16 & $-19538.5$ & $\textbf{93.61} \pm 2.85$ & $24.52 \pm 21.22$ \\
MedicalDense\_20 & $-32023.8$ & $\textbf{93.71} \pm 6.25$ & $23.94 \pm 3.6$ \\
\bottomrule
\end{tabular}
\end{table}

\subsection{Additional QAOA results under time-matched budgets}
\label{sec:time_matched}

Table~\ref{tab:time_matched_results} reports the time-matched QAOA results on four representative benchmarks (MaxCut\_12, MaxCut\_20, Knapsack\_16, MedicalDense\_20). We evaluate SPIQ-T, a time-matched variant of SPIQ~\cite{bharadwaj2026spiq}, under a wall-time budget equal to the average time CRiSP took to reach its best solution. Under this equal-time comparison, SPIQ-T reaches 52.27\%, 40.63\%, 24.52\% and 23.94\% average accuracy on MaxCut\_12, MaxCut\_20, Knapsack\_16 and MedicalDense\_20, respectively, while CRiSP achieves 100\%, 53.99\%, 93.61\% and 93.71\% average accuracy. This indicates that CRiSP's advantage comes from a more effective search within the Clifford space, not merely from spending more classical compute.

\begin{figure}[t]
    \centering
    \includegraphics[width=0.6\textwidth]{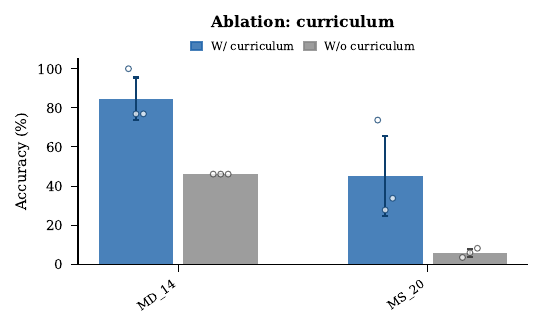}
    \caption{Ablation study with and without incremental horizon curriculum, showing CRiSP accuracy. The bar indicates the average accuracy, the error bar displays std, and the points correspond to results from three independent runs. }
    \label{fig:abl_curriculum}
\end{figure}

\subsection{Ablation study on incremental horizon curriculum}
\label{sec:abl_curriculum}
To examine the contribution of the incremental horizon curriculum, we test CRiSP with and without the curriculum on two representative benchmarks where the total episode length is large and the energy landscape is highly barren, namely MedicalDense\_14 and MedicalSparse\_20. Due to the substantial computational cost of full-horizon training on MedicalDense tasks, the ablation is conducted under a reduced setting with $3{,}000$ training episodes, a simulation count of $30$ for both the warmup and standard phases, and a learning rate of $3\times 10^{-4}$, while keeping all other hyperparameters the same as the configuration for the main results. Both variants are evaluated under this same setting to ensure a fair comparison. The ablation for MedicalSparse\_20 is performed using the identical experimental setup as the main experiments.

As shown in Fig.~\ref{fig:abl_curriculum}, without the curriculum, CRiSP struggles to effectively explore the full search space within the allocated training budget, causing substantial performance degradation on both benchmarks. On MedicalDense\_14, the curriculum pushes the average accuracy from $46.14\%$ to $84.59\%$. This indicates that the gradual expansion of the search horizon enables the agent to uncover partially promising configurations in the early stages of training, and to further focus its exploration on those regions as the horizon extends toward its full length. The performance gap becomes even more drastic on MedicalSparse\_20, where the reward landscape is extremely sparse. Here, the mean accuracy improves $7.62\times$, from $5.92\%$ to $45.11\%$, because the curriculum allows the agent to find positive reward signals during early episodes within confined search spaces---regions that would otherwise remain unexplored during full-horizon exploration. These results confirm that the incremental horizon curriculum is highly beneficial in addressing tasks with large parameter counts and challenging energy landscapes.

\subsection{Ablation study for CRiSP's components}
\label{sec:abl_component}
To examine which components drive CRiSP's gains, we conduct an ablation on MaxCut\_12 under a matched stabilizer simulation budget.
We evaluate variants that remove CRiSP's best-game buffer, temperature decay, or Dirichlet noise, as well as variants that replace its Transformer-based policy with a simpler two-layer linear architecture, or that remove policy guidance entirely via unguided MCTS or random search.
As shown in Table~\ref{tab:abl_crisp_component}, ablating any of these factors leads to a clear degradation in performance compared to the full model (100\% $\pm$ 0\% average accuracy). 
This confirms that each component meaningfully contributes to CRiSP's search capability.

\begin{table}[h]
\centering
\caption{Ablation comparing CRiSP with variants that remove the best-game buffer, temperature decay, or Dirichlet noise, or that replace its Transformer-based policy with a simpler policy architecture (two linear layers), or that remove policy guidance entirely via unguided MCTS or random search. Evaluated on MaxCut\_12 ($E_{\text{opt}}=-88$), averaged over three seeds (mean $\pm$ std), under the same total evaluation budget as CRiSP's training (60,000 stabilizer evaluations).}
\label{tab:abl_crisp_component}
\footnotesize
\begin{tabular}{l c c}
\toprule
 & Accuracy & Avg. Energy\\
\midrule
 \textbf{CRiSP (Ours)} & \textbf{100\% $\pm$ 0\%} & \textbf{-88} \\
\midrule
 Unguided MCTS & $85.23\% \pm 4.95\%$ & -75 \\
\midrule
 No Dirichlet noise & $83.33\% \pm 17.69\%$ & -73.33 \\
\midrule
 No temperature decay & $79.17\% \pm 4.59\%$ & -69.67 \\
\midrule
 Simpler policy & $77.27\% \pm 7.10\%$ & -68 \\
\midrule
 No best-game buffer & $70.83\% \pm 11.04\%$ & -62.33 \\
\midrule
 Random search & $38.26\% \pm 0.66\%$ & -33.67 \\

\bottomrule
\end{tabular}
\end{table}

\section{Quantum computing background}

\subsection{Qubits, quantum gates, and the Clifford group}
\label{subsec:qubits_gates_cliffs}

At the core of quantum computation lies the quantum bit, or \emph{qubit}. In stark contrast to deterministic classical bits, a single qubit can occupy a linear combination, or superposition, of its computational basis states, denoted mathematically as $\ket{\psi} = \alpha \ket{0} + \beta \ket{1}$. This principle scales exponentially: the state of an $N$-qubit register is represented by a joint superposition across all $2^N$ possible classical bitstrings within $\qty{0,1}^N$. To evolve these quantum states, one applies continuous unitary transformations referred to as \emph{quantum gates}. Theoretical and experimental focus is frequently directed toward single- and two-qubit operations, as they collectively provide a universal basis capable of executing any arbitrary quantum algorithm \cite{Nielsen_Chuang_2010}.

Within quantum information theory, the single-qubit Pauli operators, denoted as $\PG = \qty{I, X, Y, Z}$, serve as a foundational toolkit. The identity matrix $I$ acts trivially by preserving the state, whereas the $X$ operator induces a direct bit flip ($\ket{0} \leftrightarrow \ket{1}$). Conversely, the $Z$ operator functions as a phase flip, appending a negative sign strictly to the excited state ($Z\ket{1} = -\ket{1}$). The $Y$ gate effectively couples both bit and phase flips, defined as $Y = iXZ$. To describe multi-qubit interactions, the $N$-qubit Pauli group $\PG_N$ is constructed via the $N$-fold tensor product of these fundamental single-qubit operators, such that $\PG_N = \PG^{\otimes N}$.

Expanding on this foundation yields the \emph{Clifford group} $\CG_N$, a paramount algebraic structure in quantum computing. For an $N$-qubit system, $\CG_N$ encompasses all quantum operations $C$ that act as normalizers for the Pauli group $\PG_N$. This requirement dictates that mapping a Pauli operator through a Clifford gate must return another valid Pauli operator, expressed mathematically as:
\begin{equation}
    \label{eq:clifford_group}
    C P C^\dagger \in \pm \PG_N \quad \forall \, P \in \PG_N.
\end{equation}
The transformation $C P C^\dagger$ is formally known as the \emph{conjugation} of a Pauli matrix $P$ by a Clifford operation $C$, which maps to a new Pauli string, potentially with an inverted sign. Because Pauli operators trivially normalize themselves, the Pauli group inherently exists as a strict subgroup within the broader Clifford group ($\PG_N \subset \CG_N$).

\subsection{Universal quantum gate sets and associated costs}
\label{subsec:universal_gate_set}
To perform arbitrary quantum computations, a universal gate set is necessary to approximate any unitary operator to a desired precision. In the context of fault-tolerant quantum computing (FTQC), the discrete Clifford+T set is a standard choice. By themselves, Clifford operations—such as the Pauli, Hadamard, and CNOT gates—are non-universal. However, supplementing them with the non-Clifford T gate unlocks universality, allowing for the precise approximation of continuous operations \cite{kliuchnikov2012fast}. Tools like Gridsynth \cite{ross2016gridsynth} are used to compile continuous arbitrary gates into discrete Clifford+T sequences. The drawback is that achieving higher mathematical precision necessitates exponentially longer gate sequences, leading to a massive computational overhead that is especially prohibitive in the early FTQC era.

This bottleneck is a direct consequence of the Eastin-Knill theorem \cite{eastin2009restrictions}, which establishes that no quantum error correction (QEC) code can natively and fault-tolerantly implement a universal gate set. As a result, standard QEC schemes like the surface code can easily protect Clifford operations but fundamentally cannot correct errors for non-Clifford gates such as the T gate or continuous $R_z(\theta)$ rotations.

To execute these non-Clifford operations, FTQC architectures rely on a resource-intensive procedure known as \textit{magic state distillation} (or T state distillation) \cite{ bravyi2005universal, litinski2019magic}. This method generates high-fidelity resource states, such as the T state $\ket{T} = (1/\sqrt{2})(\ket{0} + \exp(i\frac{\pi}{4})\ket{1})$. Once prepared, these low-error T states are consumed alongside an arbitrary data qubit to successfully apply a T gate to the circuit, though this process incurs substantial qubit-cycle penalties.

The physical implementation of this process occurs in "magic state factories" \cite{litinski2019magic}. The procedure begins by initializing several surface code patches in the $\ket{+} = (\ket{0} + \ket{1})/\sqrt{2}$ state and applying a series of logical rotations. A designated subset of these qubits is then measured to extract an error syndrome. If this signature flags an error during the logical rotations, the entire batch is discarded, and the protocol restarts. If the signature is clear, the remaining unmeasured qubits yield distilled T states. Because the syndrome measurements cannot catch every possible error type, the final T states are of exceptionally high fidelity, but not strictly perfect.

The footprint of a T state distillation protocol is largely defined by its input-to-output patch ratio, alongside the spatial and temporal code distances ($d_X$, $d_Z$, and $d_m$). For example, a standard (15-to-1)\(_{7, 3, 3}\) factory consumes 15 raw input patches to generate a single distilled T state, utilizing distances of $d_X = 7$, $d_Z = 3$, and $d_m = 3$. Suppressing the error rate of the output T states requires scaling up these dimensions, which directly inflates the space-time volume of the factory. Consequently, protocols designed for higher output yields—such as a (20-to-4) factory—demand significantly greater hardware resources than a basic (15-to-1) configuration operating under the same code distances.

\subsection{Achieving utility with VQAs}\label{sec:vqa}

Although near-term quantum platforms are inherently noisy, specific algorithmic frameworks—most notably variational quantum algorithms (VQAs)—offer a potential pathway to surpassing classical computational capabilities. At its core, a VQA evaluates a parameterized quantum circuit (the ansatz) with respect to a problem-specific Hermitian operator (the Hamiltonian) to compute a scalar objective function, typically the system's energy. This energy landscape is then navigated by classical optimization routines, such as Nelder-Mead or SPSA \cite{TILLY20221}, which iteratively update the circuit parameters to locate the minimum. A key theoretical advantage of VQAs is their Optimal Parameter Resilience (OPR) \cite{wang2021can}, a property suggesting that the optimal parameter configurations discovered under noisy hardware conditions generally align with the ideal, noiseless solutions. In practice, however, the sheer magnitude of noise in the Noisy Intermediate-Scale Quantum (NISQ) era frequently overwhelms this inherent robustness, necessitating explicit error reduction techniques to solve complex, practical problems.

Historically, efforts to suppress errors on near-term hardware have heavily favored quantum error mitigation. Although a vast array of mitigation protocols has been developed for NISQ architectures \cite{dangwal2023varsaw, ravi2021vaqem, ravi2023cafqa, ravi2022qismet, li2022paulihedral, jin2024tetris, seifert2024clapton, giurgica2020digital}, these techniques alone have proven insufficient to bridge the gap to practical utility. To genuinely surpass classical computational capabilities, many VQA applications demand extremely high operational fidelities—often exceeding $99\%$—which mitigation alone struggles to deliver under severe hardware noise. Consequently, while advancing novel error mitigation strategies remains important, there is a growing consensus that transitioning toward early, lightweight, or partial implementations of quantum error correction is an unavoidable prerequisite for sufficiently suppressing noise and realizing true quantum advantage.

\subsection{Solving combinatorial optimization problems with QAOA}
\label{subsec:qaoa_problem_derivation}

Combinatorial optimization fundamentally involves identifying the optimal configuration from an exponentially large discrete search space. This paradigm underpins critical applications across logistics \cite{logistics_prob}, finance \cite{finance_prob}, and operations research \cite{operations_prob}. Because these problems are overwhelmingly NP-hard, finding exact classical solutions becomes computationally intractable as system sizes scale. In this context, problem formulation plays a pivotal role. The most common formulation is Quadratic Unconstrained Binary Optimization (QUBO), defined as $\min x^T Q x$ where $x \in \{0,1\}^n$ and $Q \in \mathbb{R}^{n \times n}$ is symmetric. This can be extended to Polynomial Unconstrained Binary Optimization (PUBO) to capture higher-order interactions ($\min \sum_i a_i x_i + \sum_{i<j} b_{ij} x_i x_j + \sum_{i<j<k} c_{ijk} x_i x_j x_k + \cdots$), and further generalized to Polynomial Constrained Binary Optimization (PCBO) by incorporating constraints. While PCBO models offer greater expressivity, they introduce significant computational complexity. 

Classical exact solvers like CPLEX and Gurobi \cite{cplex, gurobi} guarantee optimality through branch-and-bound and relaxation methods, but they suffer from exponential scaling. Conversely, heuristic approaches like simulated annealing and genetic algorithms \cite{boros2007local} provide approximate scalability for larger problem instances. Often, PCBO structures are quadratized into standard QUBO forms \cite{boros2002quadratization, anthony2017quadratization} to leverage existing solvers, though this comes at the expense of introducing auxiliary variables. Consequently, practically scaling these solutions continues to depend heavily on heuristic and approximation frameworks \cite{kochenberger2014unconstrained, glover2019qubo}.

To address these scaling limitations computationally, the Quantum Approximate Optimization Algorithm (QAOA) \cite{farhi2014quantum} offers a hybrid quantum-classical approach tailored for optimization tasks. A standard QAOA circuit is constructed by alternately applying $p$ layers of two parameterized unitaries: a problem-specific cost operator $U(\gamma, H_C) = e^{-i H_C \gamma}$ and a mixing operator $U(\beta, H_M) = e^{-i H_M \beta}$. Starting from an initial uniform superposition of all computational basis states, $\ket{s} = \frac{1}{\sqrt{2^n}} \sum_{z} \ket{z}$, the system is variationally evolved for $p$ repetitions to approximate the low-energy configurations of the target cost Hamiltonian $H_C$. 

Unlike the Variational Quantum Eigensolver (VQE), which typically targets physical ground states, QAOA is specifically designed to minimize classical discrete cost functions. However, because the standard QAOA ansatz restricts expressivity through its rigid, problem-specific layered structure, navigating its non-convex, instance-dependent optimization landscape is notoriously difficult \cite{wang18, farhi22, barak15, hastings19, marwaha21, chou22, lin16, farhi15, hadfield19, streif20}. While increasing the circuit depth ($p$) theoretically improves expressivity, it concurrently amplifies noise susceptibility on near-term hardware. 

An alternative strategy to enhance expressivity without deepening the ansatz is to expand the classical parameter space per layer via multi-angle QAOA (ma-QAOA) \cite{herrman2022maqaoa}. Rather than tying all operations in a layer to a single pair of variables, ma-QAOA assigns independent variational angles to each distinct term within the cost and mixer Hamiltonians:
\begin{align}
    U(H_C, \vec{\gamma_\ell}) &= e^{-i \sum_{a=1}^{m} H_{C,a} \gamma_{\ell,a}} = \prod_{a=1}^{m} e^{-i H_{C,a} \gamma_{\ell,a}}, \\
    U(H_M, \vec{\beta_\ell}) &= e^{-i \sum_{b=1}^{n} H_{M,b} \beta_{\ell,b}} = \prod_{b=1}^{n} e^{-i H_{M,b} \beta_{\ell,b}}.
\end{align}
For a system with $m$ clauses and $n$ qubits, this expands the parameter count from the standard $2p$ to $(m+n)p$. Standard QAOA naturally emerges as a special case of this formulation, whereas ma-QAOA allows classical optimizers to explore a vastly richer, multidimensional solution landscape.

Despite this increased expressivity, QAOA remains highly sensitive to parameter initialization \cite{Sack_2021, blekos2024review}. Poor starting configurations frequently trap the optimizer in suboptimal local minima or barren plateaus lacking useful gradient information \cite{lee_2022}. Existing initialization heuristics, such as INTERP \cite{INTERP} or TQA \cite{Sack_2021}, are typically customized for specific problem structures and struggle to generalize or scale seamlessly. This lack of robust, universally applicable initialization remains a major barrier to deploying QAOA for complex, real-world tasks.

Once a viable starting point is established, the continuous optimization loop updates the parameters within the interval $[-\pi, \pi)$—extending the search well beyond discrete Clifford angles. Depending on the landscape's smoothness and the presence of hardware noise, this loop may employ gradient-free algorithms like COBYLA \cite{Zhang_2023}, SPSA \cite{SPSA}, or Nelder-Mead \cite{Nelder-mead}, or gradient-based methods like L-BFGS-B \cite{lbfgsb} and Adam \cite{adam-optimizer}. This iterative refinement proceeds until the algorithm converges or triggers a predefined iteration limit to conserve quantum resources.

Given the extreme non-convexity of the QAOA landscape, a single optimization trajectory—even one initialized with a high-quality warm-start—carries a substantial risk of converging to an unfavorable local minimum \cite{Sack_2021, lee2024iterative}. To mitigate this vulnerability, multi-start optimization strategies \cite{rinnooy1987stochasticI} are widely employed, launching several independent optimization routines from a diverse collection of initial seeds to ensure robust convergence. 

To empirically assess our work in the QAOA setting, we select a diverse set of optimization tasks. First, we leverage MaxCut as our primary QUBO benchmark. For a graph \( G \) with edge set \( E(G) \), the corresponding cost Hamiltonian is defined as \( H_{\text{C}} = \sum_{(i,j) \in E(G)} \frac{1}{2} w_{ij} \left( I - Z_i Z_j \right) \), where \( w_{ij} \) denotes the weight of edge \( (i,j) \), and \( Z_i \) and \( Z_j \) are Pauli-\(Z\) operators acting on the qubits associated with vertices \( i \) and \( j \), respectively. MaxCut serves as an effective benchmark for evaluating our framework, enabling systematic analysis across varied graph topologies under a unified optimization setting. Specifically, we evaluate on weighted complete graphs, which exhibit dense connectivity, leading to highly entangled and globally coupled optimization landscapes.

For PCBO evaluation, we consider the Knapsack problem and a biomedical high-order interaction task~\cite{cancer_genome} as representative benchmarks. The Knapsack problem is mapped to a QUBO Hamiltonian using the Qiskit Optimization module~\cite{qiskit23}, where constraint satisfaction (e.g., capacity limits) is enforced through penalty terms. Problem instances are generated with randomized item weights and values to ensure a range of nontrivial optimization scenarios.

For the biomedical benchmark~\cite{cancer_genome}, each binary variable corresponds to a gene-associated feature, with bit assignments indicating selection within a subset. The PCBO representation captures both individual feature relevance and higher-order statistical dependencies—including pairwise and triple-wise mutual information—modeled as a weighted hypergraph. The objective is to select \( M \) features from a total of \( N \) candidates to maximize predictive relevance while reducing redundancy. This selection constraint is imposed via a Hamming-weight condition, enforced through an additional Lagrangian penalty term. 
In our experiments, we set \( M=4 \) for the MedicalDense tasks, \( M=8 \) for the 15-qubit MedicalSparse tasks, and \( M=10 \) for the MedicalSparse\_20 task.
Further details regarding these QAOA benchmarks can be found in~\cite{bharadwaj2026spiq}.

\subsection{Solving scientific problems with VQE}
\label{subsec:vqe_problem_derivation}

A prominent implementation of the VQA paradigm is the Variational Quantum Eigensolver (VQE) \cite{TILLY20221}, which is explicitly designed to estimate the ground-state energy—or the minimum eigenvalue—of a target Hamiltonian $\ham$. VQE has become a cornerstone algorithm in computational physics and chemistry, routinely applied to compute molecular dissociation curves as a function of bond length or to resolve the minimum energy configurations of quantum spin models. Because of its practical importance, optimizing the reliable execution of VQE has been the subject of extensive architectural and systems-level research \cite{ravi2023cafqa, adaptvqe, wang2022quantumnas, ravi2021vaqem}. In this study, we leverage VQE to investigate fundamental physical systems, detailing the specific target models below.

Within the realm of statistical mechanics, the Ising \cite{ising} and Heisenberg \cite{heisenberg} models serve as foundational frameworks for describing interacting magnetic spin systems. These models are renowned for hosting rich quantum phenomena, including dimension-dependent phase transitions and highly entangled, non-classical ground states. Our analysis specifically targets one-dimensional configurations governed by uniform coupling parameters. The first is the one-dimensional transverse-field \emph{Ising model}, defined by the Hamiltonian:
\begin{equation}
    \ham = J \sum_{i=1}^{N-1} X_i X_{i+1} + \sum_{i=1}^N Z_i.
\end{equation}
This operator models a chain of $N$ coupled spins (qubits) where nearest-neighbor interactions along the $x$-axis are governed by a coupling strength $J$, while each spin simultaneously interacts with a uniform external magnetic field of unit strength aligned along the $z$-axis.

Additionally, we examine a restricted variant of the field-free Heisenberg model, which generalizes spin-spin coupling across all spatial dimensions. Specifically, we focus on the Hamiltonian:
\begin{equation}
    \label{eq:xxz}
    \ham = \sum_{i=1}^{N-1} \qty(J X_i X_{i+1} + J Y_i Y_{i+1} + Z_i Z_{i+1}).
\end{equation}
In this formulation, the system's energy scale is normalized by fixing the $ZZ$-interaction at unit strength. Because the transverse interactions along both the $X$ and $Y$ axes are constrained to share an identical coupling constant $J$, this specific parameterization is conventionally identified as the \emph{XXZ model}.

\section{Numerical results for QAOA and VQE benchmarks}
\label{sec:numerical_results}

Table~\ref{tab:qaoa_results} shows the numerical results for QAOA benchmarks, and the simulation outcomes for VQE tasks are provided in Table~\ref{tab:vqe_results}.

\begin{table}[h]
\centering
\caption{
    QAOA benchmark results. 
    $n$ and $N_\text{params}$ denote the number of qubits and circuit parameters, respectively.
    $E_\text{opt}$ represents the ground state energy of the problem.
    CRiSP and SPIQ are compared under equal circuit evaluation budgets.
    SPIQ-R denotes SPIQ evaluated with the same number of update rounds as CRiSP.
    Results are reported as mean $\pm$ standard deviation (std) across three independent runs.
    $^\dagger$ indicates tasks where the incremental horizon curriculum is applied.
    Bold values denote the best average performance for each task.
    MC: MaxCut, Kn: Knapsack, MD: MedicalDense, MS: MedicalSparse.
}
\label{tab:qaoa_results}
\resizebox{\textwidth}{!}{
\begin{tabular}{llrrrrrr}
\toprule
Category & Task & $n$ & $N_\text{params}$ & $E_\text{opt}$ & CRiSP (Ours) (\%) & SPIQ (\%) & SPIQ-R (\%) \\
\midrule
\multirow{4}{*}{MaxCut}
 & MC\_8  & 8  & 36  & $-46$       & $\textbf{100} \pm 0$ & $83.33 \pm 16.02$ & $86.96 \pm 18.95$ \\
 & MC\_12 & 12  & 78  & $-88$       & $\textbf{100} \pm 0$   & $47.35 \pm 9.46$  & $51.89 \pm 11.44$ \\
 & MC\_16 & 16  &  136 & $-155$      & $\textbf{80.22} \pm 0.74$   & $33.55 \pm 6.15$  & $37.85 \pm 9.51$  \\
 & MC\_20 & 20  & 210  & $-192$      & $\textbf{53.99} \pm 6.1$  & $4.69 \pm 4.45$   & $9.55 \pm 4.18$   \\
\midrule

\multirow{4}{*}{Knapsack}
 & Kn\_4  & 8  & 44  & $-255$      & $\textbf{99.74} \pm 0.45$   & $91.63 \pm 2.27$  & $91.63 \pm 2.27$  \\
 & Kn\_9  & 14  & 119  & $-2309.5$   & $\textbf{94.56} \pm 4.15$   & $76.24 \pm 19.86$ & $81.50 \pm 16.75$ \\
 & Kn\_12 & 18  & 189  & $-7079.5$   & $\textbf{92.18} \pm 6.59$   & $15.34 \pm 12.83$ & $29.68 \pm 4.87$  \\
 & Kn\_16 & 22  & 275  & $-19538.5$  & $\textbf{93.61} \pm 2.85$   & $3.6 \pm 6.23$   & $4.26 \pm 7.2$   \\
\midrule

\multirow{3}{*}{MedicalDense}
 & MD\_12$^\dagger$ & 12  & 310  & $-24013.6$  & $\textbf{99.98} \pm 0.01$   & $73.58 \pm 6.36$  & $86.08 \pm 10.49$ \\
 & MD\_14$^\dagger$ & 14  & 483  & $-26014.2$  & $\textbf{96.14} \pm 3.84$   & $37.16 \pm 4.44$  & $47.41 \pm 8.88$  \\
 & MD\_20$^\dagger$ & 20  & 1370  & $-32023.8$  & $\textbf{93.71} \pm 6.25$   & $2.08 \pm 3.6$   & $4.16 \pm 3.6$   \\
\midrule

\multirow{3}{*}{MedicalSparse}
 & MS\_15\_1$^\dagger$ & 15  & 83  & $-0.365$    & $\textbf{38.81} \pm 45.02$  & $21.87 \pm 19.75$ & $24.91 \pm 23.19$ \\
 & MS\_15\_2$^\dagger$ & 15  & 84  & $-0.380$    & $\textbf{83.74} \pm 22.56$  & $49.82 \pm 7.31$  & $51.38 \pm 4.74$  \\
 & MS\_20$^\dagger$    & 20  & 107  & $-0.168$    & $\textbf{45.11} \pm 24.96$  & $27.17 \pm 23.11$ & $34.87 \pm 17.46$ \\
\midrule
\multicolumn{5}{r}{\textbf{Overall Mean}} & $\mathbf{83.7}$ & $40.53$ & $45.87$ \\
\bottomrule
\end{tabular}
}
\end{table}

\begin{table}[h]
\centering
\caption{
    VQE results on Ising and XXZ models.
    CRiSP and CAFQA are compared under equal circuit evaluation budgets.
    Results are reported as mean $\pm$ std across three independent runs.
    Bold values indicate the best average performance for each task.
}
\label{tab:vqe_results}
\begin{tabular}{llrrrrrr}
\toprule
Model & Task & $n$ & $N_\text{param}$ & $E_\text{opt}$ & CRiSP (Ours) (\%) & CAFQA (\%) \\
\midrule
\multirow{3}{*}{Ising}
 & Ising\_0.5 & 10 & 40 & $-10.570$ & $\textbf{94.61} \pm 0.00$ & $80.42 \pm 0.00$ \\
 & Ising\_1.0 & 10 & 40 & $-12.381$ & $\textbf{75.38} \pm 4.66$ & $\textbf{75.38} \pm 9.33$ \\
 & Ising\_2.0 & 10 & 40 & $-19.531$ & $\textbf{92.16} \pm 0.00$ & $71.68 \pm 0.00$ \\
\midrule
\multirow{3}{*}{XXZ}
 & XXZ\_0.5 & 10 & 40 & $-11.665$ & $\textbf{77.15} \pm 0.00$ & $74.29 \pm 4.95$ \\
  & XXZ\_1.0 & 10 & 40 & $-17.032$ & $\textbf{74.37} \pm$ $3.39$ & $72.41 \pm 3.39$ \\
 & XXZ\_2.0 & 10 & 40 & $-28.722$ & $\textbf{75.44} \pm 10.64$ & $69.63 \pm 0.00$ \\
\midrule
\multicolumn{5}{r}{\textbf{Overall Mean}} & $\mathbf{77.93}$ & $73.97$ \\
\bottomrule
\end{tabular}
\end{table}

\end{document}